\documentclass[USenglish, twocolumn, final]{article}

\usepackage[utf8]{inputenc}%(only for the pdftex engine)
\usepackage{microtype}
\usepackage[T1]{fontenc} % Necessary to avoid error "pdfTex error (font expansion): ..."
\usepackage{lmodern}     % Necessary to avoid error "pdfTex error (font expansion): ..."
\usepackage[square, numbers, sort&compress]{natbib}
\usepackage{subfig}
\usepackage{tikz}
\usepackage{pgfplots}
\usepackage{mathtools}

\usepackage{amsfonts}
\usepackage{amsmath}
\usepackage[left=0.68in, right=0.68in, bottom=1.225in, top=0.825in]{geometry}

\newtheorem{remark} {Remark}

\newcommand{\R}{\mathbb R}

\DeclareMathOperator{\diag}{diag}
\DeclareMathOperator{\blkdiag}{blkdiag}

\usetikzlibrary{calc}

\newcommand{\outputerrorshiftplot}[2]{
\begin{tikzpicture}
\begin{axis}[heatmap_style,
	xtick={#2},
	xticklabel={\pgfmathparse{int(\tick+1)}\pgfmathresult},
	]

\addplot[matrix plot*, unbounded coords=jump]
table[
x index=0,
y index=1,
meta index=3,
meta expr={log10(\thisrowno{3})},
]
{./results/shifts/output_error_shifts_#1.dat};
\end{axis}
\end{tikzpicture}
}
\usetikzlibrary{calc}

\newcommand{\plotcomparisonwithpod}[7]{
\begin{tikzpicture}
\pgfmathsetmacro\xticktwo{2*#6}
\pgfmathsetmacro\xtickfour{4*#6}
\begin{axis}[plot_style,
	xmin=4,
	xmax={\xtickfour},
	xlabel={{Number of reduced states, $n_r$}},
	ylabel={{#7}},
	xtick/.expanded={#6, \xticktwo, ..., \xtickfour},
	ymode=log,]

\addplot[thick, forget plot] coordinates { ({2*#6}, 1e-16) ({2*#6}, 1e16) };

\foreach \i in {1, 3, ..., #5}{
	\addplot+[ultra thick, mark=otimes] table[x index=0, y 		index=#4]{./results/comparison_with_pod/output_error_#1_#2_bt\i_#3.dat};
	\iflegendenabled\addlegendentryexpanded{BT(\i)};\fi
}

\addplot+[ultra thick, mark=otimes] table[x index=0, y index=#4]{./results/comparison_with_pod/output_error_#1_#2_pod_#3.dat};
\iflegendenabled\addlegendentry{POD};\fi
\end{axis}
\end{tikzpicture}
}

\newcommand{\outputinitialconditionplot}[2]{
	\begin{tikzpicture}
	\begin{axis}[plot_style,
		width=0.45\linewidth,
		height=0.15\textheight,
		xmin=0,
		xmax=5,
		ymin=-0.125,
		ymax= 0.05,
		xlabel={Time [s]},
		ylabel={Output [rad]},
		ytick = {-0.1, 0, 0.1},
		scaled y ticks=false,
		legend pos=south east,
		legend columns=1,
		]
	
	\addplot+[ultra thick] table[x index=0, y index=1]{./results/large_scale_example/large_scale_#1_#2.dat};
	\iflegendenabled\addlegendentry{Full order model};\fi
	
	\addplot+[ultra thick, dashed] table[x index=0, y index=2]{./results/large_scale_example/large_scale_#1_#2.dat};
	\iflegendenabled\addlegendentry{Reduced model};\fi
	\end{axis}
	\end{tikzpicture}
}

\newcommand{\outputerrorinitialconditionplot}[2]{
	\begin{tikzpicture}
	\begin{axis}[plot_style,
	width=0.45\linewidth,
	height=0.15\textheight,
	xmin=0,
	xmax=5,
	ymin=1e-6,
	ymax=1e-2,
	ymode=log,
	xlabel={Time [s]},
	ylabel={Output error [rad]},
	ytick = {1e-6, 1e-4, 1e-2},
	scaled y ticks=false
	]
	
	\addplot+[ultra thick] table[x index=0, y index=3]{./results/large_scale_example/large_scale_#1_#2.dat};
	\end{axis}
	\end{tikzpicture}
}

\newif\iflegendenabled

\pgfplotsset{compat=1.14}

\definecolor{matlab_blue}   {rgb}{   0,           0.4470,      0.7410}
\definecolor{matlab_grey}   {rgb}{   0.3,         0.3,         0.3   }
\definecolor{matlab_red}    {rgb}{   0.8500,      0.3250,      0.0980}
\definecolor{matlab_green}  {rgb}{   0.4660,      0.6740,      0.1188}
\definecolor{matlab_yellow} {rgb}{   0.9290,      0.6940,      0.1250}
\definecolor{matlab_purple} {rgb}{   0.4940,      0.1840,      0.5560}
\definecolor{matlab_teal}   {rgb}{   0.3010,      0.7450,      0.9330}
\definecolor{matlab_magenta}{rgb}{   0.6350,      0.0780,      0.1840}

\pgfplotscreateplotcyclelist{matlab_colors}{
	matlab_blue, \\
	matlab_red, \\
	matlab_green, \\
	matlab_yellow, \\
	matlab_purple, \\
	matlab_teal, \\
	matlab_magenta, \\
	matlab_grey, \\
}

\pgfplotsset{
	plot_style/.style={
		width=0.46\linewidth,
		height=0.27\linewidth,
		label style={font=\small},
		tick label style={font=\small},
		legend style={font=\scriptsize},
		ymin=1e-16,
		ymax=1e+02,
		ytick={1.0e-15, 1.0e-11, 1.0e-7, 1.0e-3, 1.0e1},
		legend pos=south west,
		legend cell align={left},
		legend columns=3,
		cycle list name=matlab_colors,
		unbounded coords=jump,
	}
}

\pgfplotsset{
	heatmap_style/.style={
		width=0.36\linewidth,
		height=0.2125\textheight,
		label style={font=\small},
		tick label style={font=\small},
		legend style={font=\small},
		xlabel = {Number of reduced states, $n_r$},
		ylabel = {shift, $\alpha$},
		enlargelimits={abs=0.5},
		point meta=explicit,
		point meta min=-13,
		point meta max=-1,
		axis on top,
		ymin=0,
		ymax=5,
		ytick={0, ..., 5},
		yticklabels={$1\mathrm{e}{-3}$,$5\mathrm{e}{-3}$, $1\mathrm{e}{-2}$,$5\mathrm{e}{-2}$, $1\mathrm{e}{-1}$,$5\mathrm{e}{-1}$, },
		colormap name=parula,
		colorbar,
		colorbar style={
			ytick={-13, -9, ..., -1},
			yticklabel=$10^{\pgfmathparse{\tick}\pgfmathprintnumber\pgfmathresult}$
		},
	}
}

\pgfplotsset
{
	colormap={blackwhite}{
		gray(0cm)=(0);
		gray(1cm)=(1)}
}

\pgfplotsset
{
	colormap={hot2}{[1cm]
		rgb255(0cm)=(0,0,0)
		rgb255(3cm)=(255,0,0)
		rgb255(6cm)=(255,255,0)
		rgb255(8cm)=(255,255,255)}
}

\pgfplotsset{
  	colormap={parula}{
  		rgb=(0.208100000000000,0.166300000000000,0.529200000000000)
  		rgb=(0.211623809523810,0.189780952380952,0.577676190476191)
  		rgb=(0.212252380952381,0.213771428571429,0.626971428571429)
  		rgb=(0.208100000000000,0.238600000000000,0.677085714285714)
  		rgb=(0.195904761904762,0.264457142857143,0.727900000000000)
  		rgb=(0.170728571428571,0.291938095238095,0.779247619047619)
  		rgb=(0.125271428571429,0.324242857142857,0.830271428571429)
  		rgb=(0.0591333333333334,0.359833333333333,0.868333333333333)
  		rgb=(0.0116952380952381,0.387509523809524,0.881957142857143)
  		rgb=(0.00595714285714286,0.408614285714286,0.882842857142857)
  		rgb=(0.0165142857142857,0.426600000000000,0.878633333333333)
  		rgb=(0.0328523809523810,0.443042857142857,0.871957142857143)
  		rgb=(0.0498142857142857,0.458571428571429,0.864057142857143)
  		rgb=(0.0629333333333333,0.473690476190476,0.855438095238095)
  		rgb=(0.0722666666666667,0.488666666666667,0.846700000000000)
  		rgb=(0.0779428571428572,0.503985714285714,0.838371428571429)
  		rgb=(0.0793476190476190,0.520023809523810,0.831180952380952)
  		rgb=(0.0749428571428571,0.537542857142857,0.826271428571429)
  		rgb=(0.0640571428571428,0.556985714285714,0.823957142857143)
  		rgb=(0.0487714285714286,0.577223809523810,0.822828571428572)
  		rgb=(0.0343428571428572,0.596580952380952,0.819852380952381)
  		rgb=(0.0265000000000000,0.613700000000000,0.813500000000000)
  		rgb=(0.0238904761904762,0.628661904761905,0.803761904761905)
  		rgb=(0.0230904761904762,0.641785714285714,0.791266666666667)
  		rgb=(0.0227714285714286,0.653485714285714,0.776757142857143)
  		rgb=(0.0266619047619048,0.664195238095238,0.760719047619048)
  		rgb=(0.0383714285714286,0.674271428571429,0.743552380952381)
  		rgb=(0.0589714285714286,0.683757142857143,0.725385714285714)
  		rgb=(0.0843000000000000,0.692833333333333,0.706166666666667)
  		rgb=(0.113295238095238,0.701500000000000,0.685857142857143)
  		rgb=(0.145271428571429,0.709757142857143,0.664628571428572)
  		rgb=(0.180133333333333,0.717657142857143,0.642433333333333)
  		rgb=(0.217828571428571,0.725042857142857,0.619261904761905)
  		rgb=(0.258642857142857,0.731714285714286,0.595428571428571)
  		rgb=(0.302171428571429,0.737604761904762,0.571185714285714)
  		rgb=(0.348166666666667,0.742433333333333,0.547266666666667)
  		rgb=(0.395257142857143,0.745900000000000,0.524442857142857)
  		rgb=(0.442009523809524,0.748080952380952,0.503314285714286)
  		rgb=(0.487123809523809,0.749061904761905,0.483976190476191)
  		rgb=(0.530028571428571,0.749114285714286,0.466114285714286)
  		rgb=(0.570857142857143,0.748519047619048,0.449390476190476)
  		rgb=(0.609852380952381,0.747314285714286,0.433685714285714)
  		rgb=(0.647300000000000,0.745600000000000,0.418800000000000)
  		rgb=(0.683419047619048,0.743476190476191,0.404433333333333)
  		rgb=(0.718409523809524,0.741133333333333,0.390476190476190)
  		rgb=(0.752485714285714,0.738400000000000,0.376814285714286)
  		rgb=(0.785842857142857,0.735566666666667,0.363271428571429)
  		rgb=(0.818504761904762,0.732733333333333,0.349790476190476)
  		rgb=(0.850657142857143,0.729900000000000,0.336028571428571)
  		rgb=(0.882433333333333,0.727433333333333,0.321700000000000)
  		rgb=(0.913933333333333,0.725785714285714,0.306276190476191)
  		rgb=(0.944957142857143,0.726114285714286,0.288642857142857)
  		rgb=(0.973895238095238,0.731395238095238,0.266647619047619)
  		rgb=(0.993771428571429,0.745457142857143,0.240347619047619)
  		rgb=(0.999042857142857,0.765314285714286,0.216414285714286)
  		rgb=(0.995533333333333,0.786057142857143,0.196652380952381)
  		rgb=(0.988000000000000,0.806600000000000,0.179366666666667)
  		rgb=(0.978857142857143,0.827142857142857,0.163314285714286)
  		rgb=(0.969700000000000,0.848138095238095,0.147452380952381)
  		rgb=(0.962585714285714,0.870514285714286,0.130900000000000)
  		rgb=(0.958871428571429,0.894900000000000,0.113242857142857)
  		rgb=(0.959823809523810,0.921833333333333,0.0948380952380953)
  		rgb=(0.966100000000000,0.951442857142857,0.0755333333333333)
  		rgb=(0.976300000000000,0.983100000000000,0.0538000000000000)
  	},
}

\begin{document}
	% ARTICLE INFORMATION
	\author{Tobias K. S. Ritschel, Frances Wei{\ss}, Manuel Baumann, and Sara Grundel}
\title{Nonlinear model reduction\\of dynamical power grid models\\using quadratization and balanced truncation}
\date{}
	
	% TITLE
	\maketitle
	
	% AFFILIATION (ONLY IN THE ARXIV VERSION)
	\renewcommand{\thefootnote}{\fnsymbol{footnote}}
\footnotetext{Tobias K. S. Ritschel, Frances Wei{\ss}, Manuel Baumann, and Sara Grundel are with Max Planck Institute for Dynamics of Complex Technical Systems, D-39106 Magdeburg, Germany. E-mails: \texttt{\{ritschel, grundel\}@mpi-magdeburg.mpg.de}.}
	
	% ABSTRACT (ONLY IN THE ARXIV VERSION)
	\begin{abstract}
		In this work, we present a nonlinear model reduction approach for reducing two commonly used nonlinear dynamical models of power grids: the \emph{effective network} (EN) model and the \emph{synchronous motor} (SM) model.
Such models are essential in real-time security assessments of power grids. However, as power grids are often large-scale, it is necessary to reduce the models in order to utilize them in real-time.
We reformulate the nonlinear power grid models as quadratic systems and reduce them using balanced truncation based on approximations of the reachability and observability Gramians.
Finally, we present examples involving numerical simulation of reduced EN and SM models of the IEEE~57 bus and IEEE~118 bus systems.
	\end{abstract}
	
	% CONTENT
	\section{Introduction}\label{sec:introduction}
Given a dynamical model, the purpose of model order reduction (MOR) is to identify another model which 1)~can be analyzed more efficiently and 2)~accurately captures the relevant dynamics and properties of the original model~\cite{morAnt05, morSchVR08}. Common analysis tasks include transient stability analysis, predictive simulation, uncertainty quantification, state estimation, and the solution of optimal control problems.

Power grids facilitate the delivery of electricity from producers to consumers. Modern power grids consist of 1)~power plants, 2)~transmission grids, 3)~distribution grids, and 4)~consumers (either industrial or residential).
In recent years, emerging technologies such as renewable energy production, charging of electric vehicles, and \emph{prosumers} have decreased the predictability of the power generation and consumption in power grids. Therefore, there is an increasing need to perform real-time grid security assessments.
However, power grid networks are often large-scale, and the commonly used dynamical power grid models are nonlinear \cite{NisM15}. Consequently, they are nontrivial to analyze in real-time.
Therefore, model reduction (also called \emph{equivalencing} \cite{JooLJ04, MilS09}) has long been used to reduce both static and dynamical models of power systems \cite{Cho13, RudPB81, WebW88}.

\emph{Balanced truncation} is a model reduction technique that involves projecting the state variables and the dynamical equations such that the states that are least affected by the inputs are also the states that affect the outputs the least. These states do not significantly affect the input-output behavior of the system and can, thus, be removed.

In fact, most model reduction techniques involve projection and truncation \cite{morAnt05}, and for linear systems, the reduced system matrices can be computed offline. However, for general nonlinear systems, the evaluation of the reduced right-hand side function will also involve the evaluation of the original right-hand side function. If this is not addressed, the reduced nonlinear model can not be analyzed significantly more efficient than the original model.

Therefore, most research on model reduction of power grid models involves reduction of a linearized system or subsystem. Researchers have used, e.g., balanced truncation \cite{AclFMetal19, Als93, CheM91, LeuKMetal19b, RamMHetal16, SanCG95, ShoA17, Stu12, StuVCetal12, StuVMetal12, StuVCetal14, ZhuGQ16}, \emph{balanced residualization} \cite{ParMA19}, \emph{Krylov methods} \cite{ChaP05, Sca15, WanYLetal13, WanYLetal15}, \emph{SVD-Krylov methods} \cite{KhaZ16}, \emph{proper orthogonal decomposition} (POD) \cite{WanLZetal14}, \emph{singular perturbation theory} \cite{ChoWPetal90, MenWZetal18, PaiA81}, variants of \emph{clustering} \cite{morCheS18, Hoc2000, WanHW06}, and sparse approximations \cite{LevB17} to reduce such linearized models.

However, nonlinear model reduction of power grid models has also been considered. Parrilo et al. \cite{ParLPetal99} use POD and Lan et al. \cite{LanZWetal16} and Zhao et al. \cite{ZhaXS13, ZhaLR17} use balanced truncation based on \emph{empirical Gramians} to reduce nonlinear models of power grids. However, they do not describe how to efficiently evaluate the reduced right-hand side function. Malik et al. \cite{MalBCetal16} use POD together with \emph{trajectory piecewise linearization} (TPWL) to reduce a nonlinear power grid model, and Purvine et al. \cite{PurCHetal17} use a clustering approach where each cluster is represented by a single generator. Osipov and Sun \cite{OsiS18} and Osipov et al. \cite{OsiDDetal18} use a hybrid approach where only a subset of the original right-hand side functions are linearized. Finally, Mlinari\'{c} et al. \cite{MliICetal18} use concepts of synchronicity to derive exact nonlinear reduced power grid models.

In this work, we use \emph{lifting} \cite{KraW19, KraW19b} to reformulate the  \emph{effective network} (EN) model and the \emph{synchronous motor} (SM) model \cite{NisM15} as quadratic models. We use a balanced truncation approach, based on approximations of the reachability and observability Gramians of the quadratic systems \cite{morBenG17, Wei19}, to reduce the quadratized models. Since the systems are quadratic, we can compute the reduced system matrices offline. Consequently, the reduced models can be analyzed more efficiently than the original models.
Finally, we present numerical examples which demonstrate the accuracy of the reduced models with numerical simulations of the IEEE 57 bus and the IEEE 118 bus systems. We use pg\_sync\_models \cite{NisM15} to obtain dynamical models of these systems.

The remainder of this paper is organized as follows. In Section \ref{sec:model}, we describe the quadratization of the EN and SM models. In Section \ref{sec:qmor}, we describe the balanced truncation approach for reducing the quadratized EN and SM models, and in Section \ref{sec:examples}, we present the numerical examples. Finally, conclusions are given in Section \ref{sec:conclusions}. % CHECK
	\section{Power system models}\label{sec:model}
The three commonly used dynamical models of power grid networks are the EN model, the SM model, and the \emph{structure-preserving} (SP) model \cite{NisM15}. All three models represent the generators and the loads (i.e., the consumers) in the power grid network as a set of coupled oscillators. The phase angle of the $i$'th oscillator (i.e., the $i$'th state variable) is described by
\begin{align}\label{eq:en:sm}
\frac{2J_i}{\omega_R} \ddot \delta_i + \frac{D_i}{\omega_R} \dot \delta_i &= F_i + f_i(\delta),
\end{align}
for $i = 1, \ldots, n_o$ where $n_o$ is the number of oscillators. Here, $\omega_R$ is a reference frequency, $H_i$ is the inertia constant, and $D_i$ is the damping constant of the $i$'th oscillator. $F_i$ is constant, and
\begin{align}\label{eq:en:sm:non:linear}
f_i(\delta) &= -\sum_{\substack{j=1\\j\neq i}}^{n_o} K_{ij} \sin\left(\delta_i - \delta_j - \gamma_{ij}\right)
\end{align}
is a nonlinear coupling term. The constant parameters $F_i$, $K_{ij}$, and $\gamma_{ij}$ depend on the steady state power flow in the network, i.e., on the solution to the \emph{power flow equations} \cite{NisM15}.
\begin{remark}
	For the EN and SM models, $J_i \neq 0$ for all $i$. However, for the SP model, $J_i = 0$ for indices $i$ representing load nodes. Consequently, the transformations described in this section would result in either 1)~a quadratic differential-algebraic system or 2)~a cubic system. Therefore, we consider only the EN and the SM models.
\end{remark}

\subsection{Transformation to first-order system}
In order to compute the Gramians in Section \ref{sec:qmor}, it is necessary to transform the second-order system \eqref{eq:en:sm} to a first-order system by augmenting the state variables with the frequencies, $\omega \vcentcolon= \dot \delta$:
\begin{subequations}\label{eq:en:sm:first:order:ode}
	\begin{align}
	\label{eq:en:sm:first:order:ode:delta}
	\dot \delta_i &= \omega_i, \\
	\label{eq:en:sm:first:order:ode:omega}
	\dot \omega_i &= -\frac{D_i}{2J_i} \omega_i + \frac{\omega_R}{2J_i} F_i + \frac{\omega_R}{2J_i} f_i(\delta),
	\end{align}
\end{subequations}
for $i = 1, \ldots, n_o$.

\subsection{Quadratization}
We further augment the state variables by introducing $s \vcentcolon= \sin(\delta)$ and $c \vcentcolon= \cos(\delta)$ and use trigonometric identities to rewrite the nonlinear function \eqref{eq:en:sm:non:linear}:
\begin{align}\label{eq:lifted:non:linear}
	f_i(s, c)
	&= -\sum_{\substack{j=1\\j\neq i}}^{n_o} K_{ij} \Big((s_i c_j - c_i s_j)\gamma_{ij}^c - (c_i c_j + s_i s_j)\gamma_{ij}^s\Big).
\end{align}
$\gamma_{ij}^s \vcentcolon= \sin(\gamma_{ij})$ and $\gamma_{ij}^c \vcentcolon= \cos(\gamma_{ij})$. The nonlinear function \eqref{eq:lifted:non:linear} is quadratic in $s$ and $c$. Furthermore, we use the chain rule to derive dynamical equations for $s$ and $c$ (which are also quadratic). The resulting lifted quadratic system is
\begin{subequations}\label{eq:en:sm:quadratic:ode}
	\begin{align}
		\label{eq:en:sm:quadratic:ode:delta}
		\dot \delta_i &= \omega_i, \\
		\label{eq:en:sm:quadratic:ode:omega}
		\dot \omega_i &= -\frac{D_i}{2J_i} \omega_i + \frac{\omega_R}{2J_i} F_i + \frac{\omega_R}{2J_i} f_i(s, c), \\
		\label{eq:en:sm:quadratic:ode:s}
		\dot s_i &= c_i \omega_i, \\
		\label{eq:en:sm:quadratic:ode:c}
		\dot c_i &= -s_i \omega_i,
	\end{align}
\end{subequations}
for $i = 1, \ldots, n_o$.
\begin{remark}\label{remark:delta}
	The right-hand sides of the lifted quadratic system \eqref{eq:en:sm:quadratic:ode} are independent of the phase angles, $\delta$.
\end{remark}

\subsubsection{Matrix form}
The quadratic system~\eqref{eq:en:sm:quadratic:ode} is in the form
\begin{align}\label{eq:quadratic:model:unshifted}
	\dot x &= Ax + H (x\otimes x) + Bu,
\end{align}
where $x \vcentcolon= \begin{bmatrix} \delta; \omega; s; c \end{bmatrix} \in\R^{4n_o}$ are the state variables, $u\in\R$ is the scalar manipulated input, and $\otimes$ denotes the Kronecker product~\cite{GolV13, HorJ91, Van00}. We use the last term in~\eqref{eq:quadratic:model:unshifted} to represent the constant terms in~\eqref{eq:en:sm:quadratic:ode}. Consequently, the (constant) manipulated inputs, $u = 1$, do not represent physical manipulable quantities.

The system matrices in~\eqref{eq:quadratic:model:unshifted} have block structure:
\begin{subequations}\label{eq:system:matrix:blocks}
	\begin{align}
	A &=
	\begin{bmatrix}
	A_{11} & \cdots & A_{14} \\
	\vdots & \ddots & \vdots \\
	A_{41} & \cdots & A_{44}
	\end{bmatrix}
	\in \R^{4n_o\times 4n_o}, \\
	H &= 
	\begin{bmatrix}
	H_{11} & \cdots & H_{14} \\
	\vdots & \ddots & \vdots \\
	H_{41} & \cdots & H_{44}
	\end{bmatrix}
	\in \R^{4n_o\times (4n_o)^2}, \\
	B &=
	\begin{bmatrix}
	B_1 \\
	\vdots \\
	B_4
	\end{bmatrix}
	\in \R^{4n_o\times 1}.
	\end{align}
\end{subequations}
$A_{ij}\in\mathbb R^{n_o\times n_o}$, $H_{ij}\in\R^{n_o\times 4n_o^2}$, and $B_i\in\mathbb R^{n_o\times 1}$.

The nonzero blocks in $A$ are
\begin{subequations}
	\begin{align}
	A_{12} &= \mathrm{I}, \\
	A_{22} &= -\frac{1}{2}J^{-1}D,
	\end{align}
\end{subequations}
where $\mathrm I$ is the identity matrix, $J = \diag \{J_i\}_{i=1}^{n_o}$, and $D = \diag \{D_i\}_{i=1}^{n_o}$.

The nonzero blocks of the Hessian matrix, $H$, are block-diagonal where each block is a row vector with $4 n_o$ elements:
\begin{subequations}
	\begin{align}
	H_{23} &= \blkdiag\left\{\begin{bmatrix} 0 & 0 & \frac{\omega_R}{2J_i} h_i^s & -\frac{\omega_R}{2J_i} h_i^c \end{bmatrix}\right\}_{i=1}^{n_o}, \\
	H_{24} &= \blkdiag\left\{\begin{bmatrix} 0 & 0 & \frac{\omega_R}{2J_i} h_i^c & \phantom{-}\frac{\omega_R}{2J_i} h_i^s \end{bmatrix}\right\}_{i=1}^{n_o}, \\
	H_{34} &= \blkdiag\left\{\begin{bmatrix} 0 & \phantom{-}e_i & 0 & 0 \end{bmatrix}\right\}_{i=1}^{n_o}, \\
	H_{43} &= \blkdiag\left\{\begin{bmatrix} 0 & -e_i & 0 & 0 \end{bmatrix}\right\}_{i=1}^{n_o}.
	\end{align}
\end{subequations}
The $i$'th element of the row vector $e_i\in\mathbb R^{n_o}$ is one, and all other elements are zero. The elements of the row vectors $h_i^s, h_i^c\in\mathbb R^{n_o}$ are
\begin{subequations}
	\begin{align}
	h_{ij}^s &=
	\begin{cases}
	K_{ij}\gamma_{ij}^s, & j \neq i, \\
	0,                           & j =    i,
	\end{cases} \\
	h_{ij}^c &= 
	\begin{cases}
	K_{ij}\gamma_{ij}^c, & j \neq i, \\
	0,                           & j = i.
	\end{cases}
	\end{align}
\end{subequations}

Finally, the nonzero block of $B$ is
\begin{align}
B_2 &= F.
\end{align}
\begin{remark}
	The Hessian matrix, $H$, is not unique. Therefore, we modify it such that it is \emph{symmetric} \cite{morBenB15}, i.e., such that $H(u\otimes v) = H(v\otimes u)$.
\end{remark}

\subsection{Nonzero initial condition}
In Section \ref{sec:qmor}, we use expressions for the Gramians of quadratic systems which require that the initial condition is zero \cite{morBenG17}. However, the state variables in the lifted quadratic model contain both sines and cosines of the phase angles. These can not simultaneously be zero. Therefore, we introduce the shifted state variables $\bar x \vcentcolon= x - x_0$ and the augmented manipulated inputs $\bar u \vcentcolon= \begin{bmatrix} u; 1 \end{bmatrix}$. $x_0$ is a given initial condition for the lifted quadratic system. Using properties of the Kronecker product \cite{HorJ91}, we derive a quadratic model for the shifted state variables (which are zero at the initial time):
\begin{align}\label{eq:shifted:quadratic:system:zero:ic}
	\dot{\bar x} &= \bar A \bar x + H (\bar x\otimes\bar x) + \bar B \bar u, & \bar x(0) &= 0.
\end{align}
In \eqref{eq:shifted:quadratic:system:zero:ic}, $\bar A = A + A_0$ and $\bar B = \begin{bmatrix} B & B_0 \end{bmatrix}$, where
\begin{subequations}
	\begin{align}
	A_0 &= H\left((\mathrm{I}\otimes x_0) + (x_0\otimes\mathrm{I})\right), \\
	B_0 &= A x_0 + H(x_0\otimes x_0).
	\end{align}
\end{subequations} % CHECK
	\section{Model reduction}\label{sec:qmor}
In this section, we describe a balanced truncation approach, based on that described by Benner and Goyal \cite{morBenG17}, for reducing the quadratic system
\begin{subequations}\label{eq:quadratic:model}
	\begin{align}
		\label{eq:quadratic:model:x}
		\dot x &= A x + H (x\otimes x) + B u, & x(0) &= 0, \\
		\label{eq:quadratic:model:y}
		y &= C x,
	\end{align}
\end{subequations}
where $x\in\R^n$, $u\in\R^m$, $y\in\R^p$, $A\in\R^{n\times n}$, $H\in\R^{n\times n^2}$, $B\in\R^{n\times m}$, and $C\in\R^{p\times n}$. \eqref{eq:quadratic:model:x} is the shifted dynamical power system model described in Section \ref{sec:model}, and \eqref{eq:quadratic:model:y} relates the outputs, $y$, to the state variables, $x$.

In order to reduce \eqref{eq:quadratic:model}, we use the projection matrices $\mathcal W, \mathcal V\in\R^{n\times n_r}$ ($\mathcal W^T \mathcal V = I$) to project and truncate the state variables ($x \approx \mathcal V \hat x$) and the dynamical equations (left multiply by $\mathcal W^T$). $n_r$ is the number of states in the reduced model. The resulting quadratic reduced order model is
\begin{subequations}
	\begin{align}
		\dot{\hat x} &= A_r \hat x + H_r (\hat x\otimes\hat x) + B_r u, \\
		\hat y &= C_r \hat x,
	\end{align}
\end{subequations}
where $\hat x\in\R^{n_r}$ and $\hat y\in\R^p$ are the reduced state variables and outputs, and $A_r\in\R^{n_r\times n_r}$, $H_r\in\R^{n_r\times n_r^2}$, $B_r\in\R^{n_r\times m}$, and $C_r\in\R^{p\times n_r}$ are the reduced system matrices given by the projections
\begin{subequations}\label{eq:reduced:model:system:matrices}
	\begin{align}
		A_r &= \mathcal W^T A \mathcal V, \\
		H_r &= \mathcal W^T H (\mathcal V\otimes \mathcal V), \\
		B_r &= \mathcal W^T B, \\
		C_r &= C \mathcal V.
	\end{align}
\end{subequations}

\begin{remark}
	The Kronecker product $\mathcal V\otimes \mathcal V\in\R^{n^2\times n_r^2}$ is prohibitively large in terms of memory requirements, even for moderately sized power grids.
\end{remark} % CHECK

\subsection{Gramians of quadratic systems}
The reachability Gramian, $P$, and the observability Gramian, $Q$, of the quadratic system \eqref{eq:quadratic:model} (with zero initial condition) are \cite{morBenG17}
\begin{subequations}\label{eq:gramian:sums}
	\begin{align}
	P &= \sum_{i=1}^\infty P_i, \\
	Q &= \sum_{i=1}^\infty Q_i,
	\end{align}
\end{subequations}
where $P_1$ and $Q_1$ satisfy the Lyapunov equations
\begin{subequations}\label{eq:generalized:lyapunov:equations:P1:Q1}
	\begin{align}
		A   P_1 + P_1 A^T + B   B^T &= 0, \\
		A^T Q_1 + Q_1 A   + C^T C   &= 0,
	\end{align}
\end{subequations}
and $P_i$ and $Q_i$ satisfy the Lyapunov equations
\begin{subequations}\label{eq:generalized:lyapunov:equations:Pi:Qi}
	\begin{align}
		A P_i + P_i A^T + H \left(\sum_{k=1}^{i-2} P_k\otimes P_{i - (k+1)}\right) H^T &= 0, \\
		A^T Q_i + Q_i A + \mathcal H^{(2)} \left(\sum_{k=1}^{i-2} P_k\otimes Q_{i - (k+1)}\right) (\mathcal H^{(2)})^T &= 0,
	\end{align}
\end{subequations}
for $i \geq 2$. $\mathcal H^{(2)}$ denotes the mode-2 \emph{matricization} of the tensor, $\mathcal H\in\R^{n\times n\times n}$, for which the mode-1 matricization, $\mathcal H^{(1)}$, is the Hessian matrix, $H$. Essentially, $\mathcal H^{(2)}$ is obtained by reordering the elements of $H$ (see Appendix~\ref{sec:matricization}).
\begin{remark}
	For even values of $i$, the solution to \eqref{eq:generalized:lyapunov:equations:Pi:Qi} is $P_i = Q_i = 0$.
\end{remark}
\begin{remark}
	Benner and Goyal \cite{morBenG17} showed that the reachability and observability Gramians \eqref{eq:gramian:sums} satisfy generalized Lyapunov equations and described an iterative scheme for solving these equations. However, the examples of power grid models considered in this work do not satisfy the convergence criteria for this scheme \cite[Theorem~5.3 on page~23]{morBenG17}. Therefore, we approximate the Gramians.
\end{remark} % CHECK

\subsection{Approximation of the Gramians}
We approximate the Gramians by truncating the sums in \eqref{eq:gramian:sums} to the first $N$ terms. Furthermore, we use low-rank approximations to reduce the memory consumption, which is a key computational bottleneck because of the Kronecker products in \eqref{eq:generalized:lyapunov:equations:Pi:Qi}. Finally, for the power system models described in Section \ref{sec:model}, some eigenvalues of $A$ are zero (whether the system is shifted to have zero initial condition or not). However, the real parts of the eigenvalues of $A$ must be striclty negative in order to guarantee the existence and uniqueness of solutions to the Lyapunov equations \eqref{eq:generalized:lyapunov:equations:P1:Q1}-\eqref{eq:generalized:lyapunov:equations:Pi:Qi}. Therefore, when computing the approximate Gramians, we replace $A$ by a shifted matrix,
\begin{align}\label{eq:shift:alpha}
	A_\alpha &= A - \alpha \mathrm{I},
\end{align}
where $\alpha\in\R$ is small and positive.

We approximate the Gramians, $P$ and $Q$, in \eqref{eq:gramian:sums}, by 1)~truncating the sums and 2)~approximating the truncated sums, i.e., we approximate $P$ by $P_T \approx \sum_{i=1}^N P_i \approx P$ and $Q$ by $Q_T \approx \sum_{i=1}^N Q_i \approx Q$. The approximations are
\begin{subequations}\label{eq:gramian:approx}
	\begin{align}
	P_T &= \tilde X_N \tilde X_N^T, \\
	Q_T &= \tilde Z_N \tilde Z_N^T,
	\end{align}
\end{subequations}
where $\tilde X_N$ and $\tilde Z_N$ are computed iteratively. $\tilde X_1 = \tilde R_1$ and $\tilde Z_1 = \tilde S_1$ where $\tilde R_1$ and $\tilde S_1$ are approximate low-rank factors of the solutions to \eqref{eq:generalized:lyapunov:equations:P1:Q1}. Next,
\begin{subequations}
	\begin{align}
	\tilde X_i &= \mathcal T_\tau\left(\begin{bmatrix} \tilde X_{i-2} & \tilde R_i\end{bmatrix}\right), \\
	\tilde Z_i &= \mathcal T_\tau\left(\begin{bmatrix} \tilde Z_{i-2} & \tilde S_i\end{bmatrix}\right),
	\end{align}
\end{subequations}
for $i = 3, 5, \ldots, N$, where $\tilde R_i$ and $\tilde S_i$ are approximate low-rank factors of the solutions to \eqref{eq:generalized:lyapunov:equations:Pi:Qi}, and $\mathcal T_\tau(\cdot)$ denotes low-rank approximation (see Appendix \ref{sec:column:compression}).

We obtain the approximate low-rank factors by
\begin{subequations}
	\begin{align}
	\tilde R_i &= \mathcal T_\tau(R_i), \\
	\tilde S_i &= \mathcal T_\tau(S_i),
	\end{align}
\end{subequations}
where $R_i$ and $S_i$ are approximations of the Cholesky factors of the solutions to \eqref{eq:generalized:lyapunov:equations:P1:Q1}-\eqref{eq:generalized:lyapunov:equations:Pi:Qi} satisfying
\begin{subequations}\label{eq:generalized:lyapunov:equations:P1:Q1:approx}
	\begin{align}
		A_\alpha   R_1 R_1^T + R_1 R_1^T A_\alpha^T + B B^T = 0, \\
		A_\alpha^T S_1 S_1^T + S_1 S_1^T A_\alpha   + C^T C = 0,
	\end{align}
\end{subequations}
and
\begin{subequations}\label{eq:generalized:lyapunov:equations:Pi:Qi:approx}
\begin{align}
	A_\alpha   R_i R_i^T + R_i R_i^T A_\alpha^T + \tilde K_{i, i-2} \tilde K_{i, i-2}^T = 0, \\
	A_\alpha^T S_i S_i^T + S_i S_i^T A_\alpha   + \tilde L_{i, i-2} \tilde L_{i, i-2}^T = 0,
\end{align}
\end{subequations}
for $i = 3, 5, \ldots, N$.

The third terms in \eqref{eq:generalized:lyapunov:equations:Pi:Qi:approx} are approximations of the third terms in \eqref{eq:generalized:lyapunov:equations:Pi:Qi}, and we compute $\tilde K_{i, i-2}$ and $\tilde L_{i, i-2}$ iteratively. $\tilde K_{i1} = \mathcal T_\tau(\Delta K_{i1})$ and $\tilde L_{i1} = \mathcal T_\tau(\Delta L_{i1})$, and
\begin{subequations}
	\begin{align}
		\tilde K_{ik} &= \mathcal T_\tau\left(\begin{bmatrix} \tilde K_{i, k-2} & \Delta K_{ik} \end{bmatrix}\right), \\
		\tilde L_{ik} &= \mathcal T_\tau\left(\begin{bmatrix} \tilde L_{i, k-2} & \Delta L_{ik} \end{bmatrix}\right),
	\end{align}
\end{subequations}
for $k = 3, 5, \ldots, i-2$, where
\begin{subequations}\label{eq:kronecker:products}
	\begin{align}
		\label{eq:kronecker:products:K}
		\Delta K_{ik} &= H \left(\tilde R_k\otimes \tilde R_{i-(k+1)}\right), \\
		\label{eq:kronecker:products:L}
		\Delta L_{ik} &= \mathcal H^{(2)} \left(\tilde R_k\otimes \tilde S_{i-(k+1)}\right).
	\end{align}
\end{subequations}
$\Delta K_{ik} \Delta K_{ik}^T$ approximates $H(P_k\otimes P_{i-(k+1)})H^T$, and $\Delta L_{ik} \Delta L_{ik}^T$ approximates $\mathcal H^{(2)}(P_k\otimes Q_{i-(k+1)})(\mathcal H^{(2)})^T$.

Finally, we use matricization to evaluate \eqref{eq:kronecker:products} efficiently (see Appendix \ref{sec:efficient:evaluation:of:kronecker:products}). % CHECK

\subsection{Balanced truncation}
Based on numerical experiments, we project and truncate $\bar \delta$, $\bar \omega$, $\bar s$, and $\bar c$ (the variables shifted to have zero initial condition), and their corresponding dynamical equations, separately. This corresponds to choosing block-diagonal projection matrices $\mathcal V$ and $\mathcal W$. Furthermore, we use the same projection matrices, $\mathcal V_\omega$ and $\mathcal W_\omega$, for all four variables:
\begin{subequations}\label{eq:projection:matrices}
	\begin{align}
		\mathcal V &= \blkdiag\{\mathcal V_\omega, \mathcal V_\omega, \mathcal V_\omega, \mathcal V_\omega\}, \\
		\mathcal W &= \blkdiag\{\mathcal W_\omega, \mathcal W_\omega, \mathcal W_\omega, \mathcal W_\omega\}.
	\end{align}
\end{subequations}

We compute the projection matrices using the square-root algorithm \cite{morAnt05} based on the second $n_o\times n_o$ diagonal blocks, $P_{\omega, T}$ and $Q_{\omega, T}$, of $P_T$ and $Q_T$ given by \eqref{eq:gramian:approx}:
%
%We compute the projection matrices using the second $n_o\times n_o$ diagonal blocks, $P_{\omega, T}$ and $Q_{\omega, T}$, of $P_T$ and $Q_T$ given by \eqref{eq:gramian:approx}, and we use the square-root algorithm \cite{morAnt05}:
%
\begin{subequations}
	\begin{align}
		\mathcal V_\omega &= R_\omega^T U_r \Sigma_r^{-1/2}, \\
		\mathcal W_\omega &= S_\omega^T V_r \Sigma_r^{-1/2}.
	\end{align}
\end{subequations}
$R_\omega$ and $S_\omega$ are the Cholesky factors of $P_{\omega, T}$ and $Q_{\omega, T}$, and $U_r$ and $V_r$ consist of the first $n_r/4$ columns of $U$ and $V$, respectively, where $R_\omega S_\omega^T = U \Sigma V^T$ is the singular value decomposition. $\Sigma_r$ is a diagonal matrix with the $n_r/4$ largest singular values on the diagonal.
\begin{remark}
	Due to numerical errors, $P_{\omega, T}$ and $Q_{\omega, T}$ may not be positive definite. In that case, we 1)~use the polar decomposition \cite{Hig86a} to compute the nearest symmetric positive semidefinite matrix \cite{Hig88} and 2)~add a small multiple
%	, e.g., $10^2\epsilon_\mathrm{mach} \|P_{\omega, T}\|_F$,
	of the identity matrix to ensure strict positive definiteness.
%	$\epsilon_\mathrm{mach}$ is the machine precision.
\end{remark} % CHECK

\subsection{Steady state adjustment}
As mentioned in Remark \ref{remark:delta}, the right-hand sides of the original nonlinear first-order EN and SM models \eqref{eq:en:sm:first:order:ode} are independent of the phase angles, $\delta$. Consequently, $\omega$, $s$, and $c$ may reach steady state regardless of the dynamics of $\delta$. Since $\dot \delta = \omega$, $\delta$ only reaches steady state if the frequencies, $\omega$, are zero in steady state. This is the case for the original model. Otherwise, \eqref{eq:en:sm:quadratic:ode:s} and \eqref{eq:en:sm:quadratic:ode:c} could not simultaneously be in steady state, as $s_i$ and $c_i$ cannot both be zero.

These aspects lead to issues in the reduced model. We explain them and their solution assuming that the initial frequencies are $\omega_0 = 0$. In that case, the reduced phase angles are given by
\begin{align}\label{eq:reduced:model:delta:hat}
	\dot{\hat \delta} = \hat \omega.
\end{align}
Analogous to the original model, the right-hand sides of the reduced model are independent of $\hat \delta$. However, the reduced frequencies, $\hat \omega$, are not necessarily zero in steady state. Consequently, $\hat \delta$ will not reach steady state. Therefore, we shift the right-hand side of \eqref{eq:reduced:model:delta:hat} by the steady state of the reduced frequencies, $\hat \omega_s$:
\begin{align}\label{eq:reduced:model:delta:hat:shifted}
	\dot{\hat \delta} = \hat \omega - \hat \omega_s.
\end{align}
Consequently, when $\hat \omega$ reaches steady state, the right-hand side of \eqref{eq:reduced:model:delta:hat:shifted} is zero, and $\hat \delta$ is also in steady state. The issues and the solution are similar if $\omega_0 \neq 0$ is used in the reduction, and we stress that we only shift the system once offline. % CHECK % CHECK

	\section{Numerical examples}\label{sec:examples}
In this section, we use numerical simulation to test the accuracy of reduced EN and SM models of the IEEE 57 bus system for different choices of parameters in the balanced truncation approach, initial conditions, and manipulated inputs. Furthermore, we demonstrate that we can effectively reduce the IEEE 118 bus system. Table~\ref{tab:numbers:of:states} shows the number of coupled oscillators in the EN and SM models, i.e., the number of states in the original second-order model \eqref{eq:en:sm}.
We use the Matlab toolboxes MATPOWER 6.0 \cite{ZimM16, ZimMT11} and pg\_sync\_models \cite{NisM15} to compute the parameters in the original model equations~\eqref{eq:en:sm}-\eqref{eq:en:sm:non:linear}.

For all tests, we use the initial conditions $\delta_0 = \omega_0 = 0$ (such that $s_0 = 0$ and $c_0 = 1$) and the manipulated inputs $u = [1; 1]$ when we reduce the models. Furthermore, the (scalar) output, $y$, is the average of the phase angles.

\begin{table}[tbh]
	\centering
	\caption{Numbers of coupled oscillators in the EN and SM models of the IEEE 57 bus and IEEE 118 bus systems.}
	\label{tab:numbers:of:states}
	\begin{tabular}{lrr}
%		\starttabularbody % Start the body of the table
							\hline
							& EN &  SM \\
							\hline %\midrule
		IEEE 57 bus system  &  7 &  57 \\
		IEEE 118 bus system & 54 & 118 \\
		\hline
	\end{tabular}
\end{table} % CHECK

\subsection{Test of the shift and number of terms}\label{sec:test:shift}
\begin{figure*}
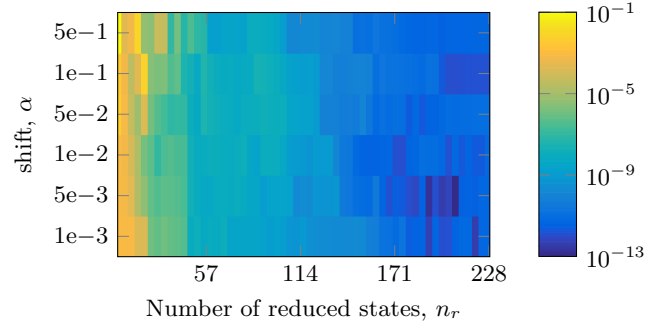
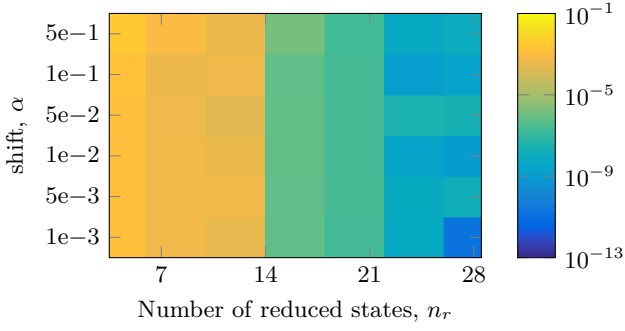
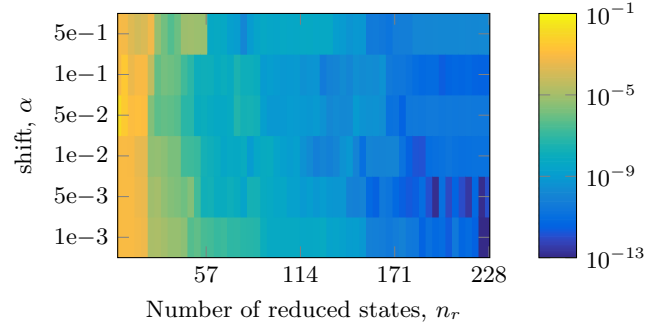
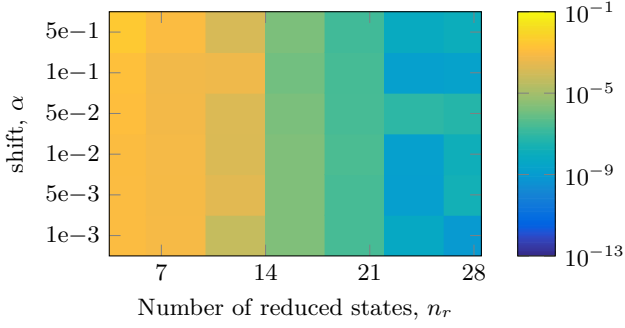
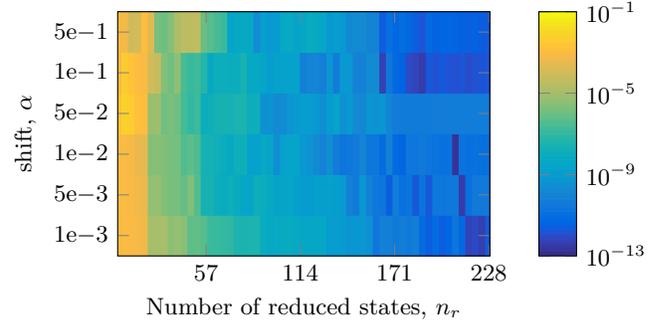

	\subfloat[The reduced EN model is obtained with $N = 1$.\label{fig:output_error_shift_en_case57_bt1}]{%
		\centering
		\outputerrorshiftplot{en_case57_bt1}{6, 13, ..., 27}
	}~ 
	\subfloat[The reduced SM model is obtained with $N = 1$.\label{fig:output_error_shift_sm_case57_bt1}]{%
		\centering
		\outputerrorshiftplot{sm_case57_bt1}{56, 113, ..., 227}
	}
	\\
	\subfloat[The reduced EN model is obtained with $N = 3$.\label{fig:output_error_shift_en_case57_bt3}]{%
		\centering
		\outputerrorshiftplot{en_case57_bt3}{6, 13, ..., 27}
	}~ 
	\subfloat[The reduced SM model is obtained with $N = 3$.\label{fig:output_error_shift_sm_case57_bt3}]{%
		\centering
		\outputerrorshiftplot{sm_case57_bt3}{56, 113, ..., 227}
	}
	\\
	\subfloat[The reduced EN model is obtained with $N = 5$.\label{fig:output_error_shift_en_case57_bt5}]{%
		\centering
		\outputerrorshiftplot{en_case57_bt5}{6, 13, ..., 27}
	}~ 
	\subfloat[The reduced SM model is obtained with $N = 5$.\label{fig:output_error_shift_sm_case57_bt5}]{%
		\centering
		\outputerrorshiftplot{sm_case57_bt5}{56, 113, ..., 227}
	}
	\\
	\subfloat[The reduced EN model is obtained with $N = 7$.\label{fig:output_error_shift_en_case57_bt7}]{%
		\centering
		\outputerrorshiftplot{en_case57_bt7}{6, 13, ..., 27}
	}~ 
	\subfloat[The reduced SM model is obtained with $N = 7$.\label{fig:output_error_shift_sm_case57_bt7}]{%
		\centering
		\outputerrorshiftplot{sm_case57_bt7}{56, 113, ..., 227}
	}
	\caption{The $L^2$-norm of the output errors for reduced EN and SM models of the IEEE 57 bus system. The reduced models are obtained using the balanced truncation approach with different 1)~shifts, $\alpha$, 2)~numbers of reduced states, $n_r$, and 3)~numbers of terms, $N$, used in approximating the Gramians.}
	\label{fig:shifts}
\end{figure*}

Fig.~\ref{fig:shifts} shows the $L^2$-norms of the output errors for reduced EN and SM models of the IEEE 57 bus system. The reduced models are obtained using the balanced truncation approach with different 1)~shifts, $\alpha$, of the $A$ matrix in \eqref{eq:shift:alpha}, 2) numbers of states in the reduced model, $n_r$, and 3) numbers of terms, $N$, in the approximate truncated sum~\eqref{eq:gramian:approx}. The simulation interval is $[0~\mathrm{s}, 2~\mathrm{s}]$.

For both the EN and the SM model, $N$ has a limited effect on the accuracy of the reduced model.
For a few combinations of $\alpha$ and $n_r$, using $N = 1$ leads to very high output errors (or even simulator breakdown, indicated by a white box).
For the SM model, and for very small $n_r$, using $N = 1$ or $N = 3$ leads to slightly lower output errors.
For $N = 3$ or higher, $\alpha$ has almost no effect on the accuracy of the reduced EN models. For the SM model, $\alpha$ slightly affects the accuracy, e.g., using $\alpha \leq 0.1$ improves the accuracy for all tested $N$.

% CREATE PLOTS
 % CHECK

\subsection{Comparison with POD}\label{sec:compare:pod}
\begin{figure*}
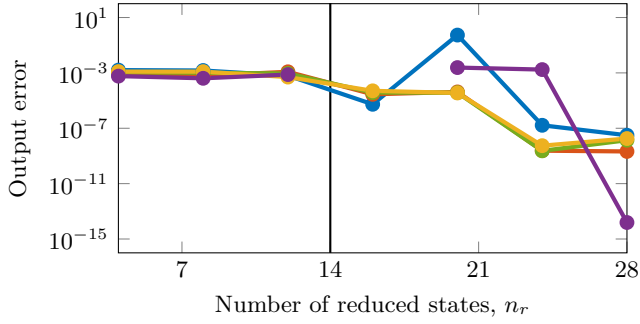
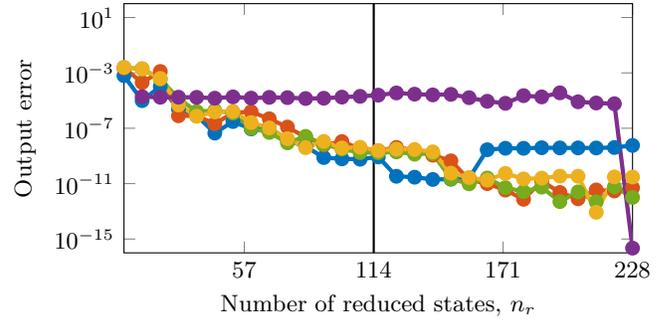
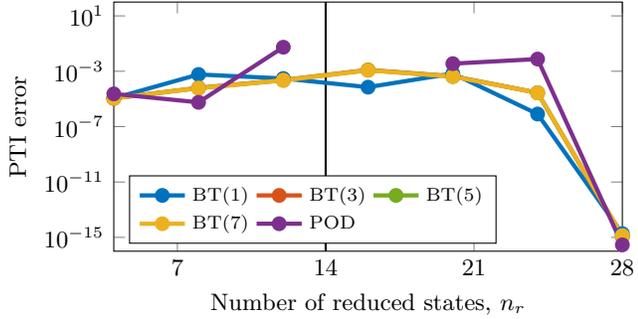
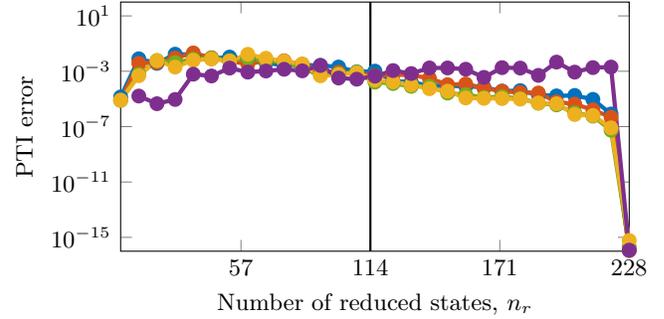

	\subfloat[Output errors for the reduced EN models.\label{fig:output_error_en_case57_x0}]{%
		\centering
		\legendenabledfalse % Enable legend
		\plotcomparisonwithpod{en}{case57}{x0_10}{1}{7}{7}{Output error} % arg. 7: 1 = L2, 2 = Linf
	}~
	\subfloat[Output errors for the reduced SM models.\label{fig:output_error:sm:case57_x0}]{%
		\centering
		\legendenabledfalse % Disable legend
		\plotcomparisonwithpod{sm}{case57}{x0_10}{1}{7}{57}{Output error} % arg. 7: 1 = L2, 2 = Linf
	}
	\\[10pt]
	\subfloat[PTI errors for the reduced EN models.\label{fig:pti_error_en_case57_x0}]{%
		\centering
		\legendenabledtrue % Disable legend
		\plotcomparisonwithpod{en}{case57}{x0_10}{3}{7}{7}{PTI error} % arg. 7: 3 = L2, 4 = Linf
	}~
	\subfloat[PTI errors for the reduced SM models.\label{fig:pti_error_sm_case57_x0}]{%
		\centering
		\legendenabledfalse % Disable legend
		\plotcomparisonwithpod{sm}{case57}{x0_10}{3}{7}{57}{PTI error} % arg. 7: 3 = L2, 4 = Linf
	}
	\caption{The $L^2$-norms of the output errors for reduced EN and SM models of the IEEE 57 bus system. In the numerical simulations, we increase the initial phase angle of the first oscillator by $0.1$~rad.}
	\label{fig:comparison:different:initial:condition}
\end{figure*}

\begin{figure*}
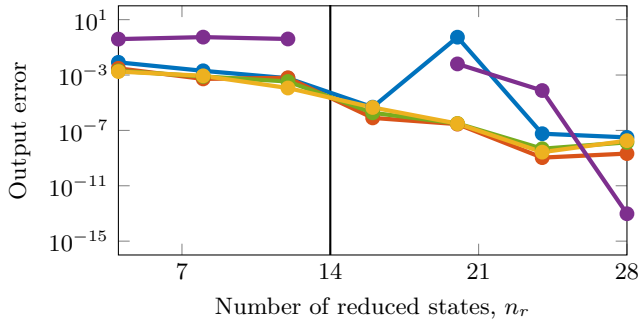
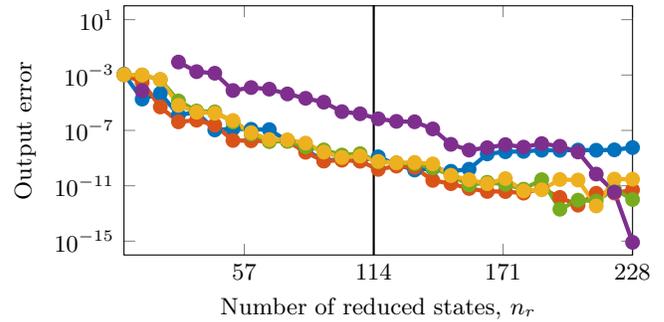
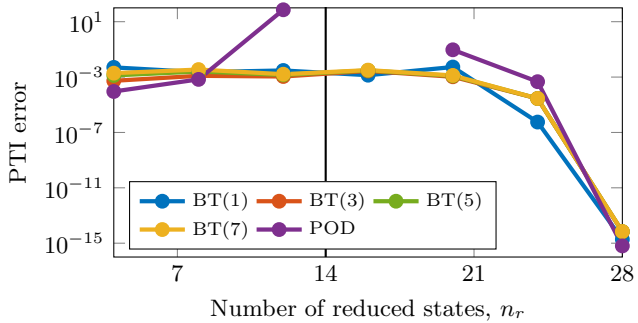
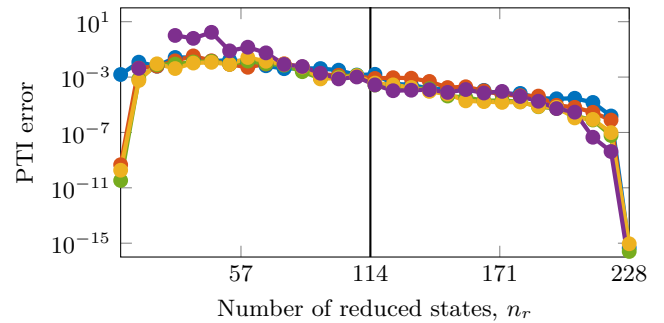

	\subfloat[Output errors for the reduced EN models.\label{fig:output_error_en_case57_u}]{%
		\centering
		\legendenabledfalse % Enable legend
		\plotcomparisonwithpod{en}{case57}{u_10}{1}{7}{7}{Output error} % arg. 7: 1 = L2, 2 = Linf
	}~
	\subfloat[Output errors for the reduced SM models.\label{fig:output_error:sm:case57_u}]{%
		\centering
		\legendenabledfalse % Disable legend
		\plotcomparisonwithpod{sm}{case57}{u_10}{1}{7}{57}{Output error} % arg. 7: 1 = L2, 2 = Linf
	}
	\\[10pt]
	\subfloat[PTI errors for the reduced EN models.\label{fig:pti_error_en_case57_u}]{%
		\centering
		\legendenabledtrue % Disable legend
		\plotcomparisonwithpod{en}{case57}{u_10}{3}{7}{7}{PTI error} % arg. 7: 3 = L2, 4 = Linf
	}~
	\subfloat[PTI errors for the reduced SM models.\label{fig:pti_error_sm_case57_u}]{%
		\centering
		\legendenabledfalse % Disable legend
		\plotcomparisonwithpod{sm}{case57}{u_10}{3}{7}{57}{PTI error} % arg. 7: 3 = L2, 4 = Linf
	}
	\caption{The $L^2$-norms of the output errors for reduced EN and SM models of the IEEE 57 bus system. In the numerical simulations, we increase the first manipulated input by $10$\%, i.e., from $u = [1; 1]$ to $u = [1.1; 1]$.}
	\label{fig:comparison:different:manipulated:inputs}
\end{figure*}

In Fig.~\ref{fig:comparison:different:initial:condition}~and~\ref{fig:comparison:different:manipulated:inputs}, we compare the balanced truncation approach with a basic POD approach~\cite[Section~9.1]{morAnt05} for reducing the EN and SM models of the IEEE 57 bus system. We apply the POD approach to the shifted quadratic system \eqref{eq:shifted:quadratic:system:zero:ic}. Therefore, as for the balanced truncation approach, the reduced system matrices can be computed offline using \eqref{eq:reduced:model:system:matrices}. We compare the $L^2$-norms of 1) the output errors and 2) the Pythagorean trigonometric identity (PTI) errors (i.e., the violation of $s_i^2 + c_i^2 = 1$).
As in Section~\ref{sec:test:shift}, we consider different numbers of terms, $N$, in the approximation of the Gramians, and the simulation interval is $[0~\mathrm{s}, 2~\mathrm{s}]$. Missing points on the graphs correspond to unsuccessful simulations.

In the first test, shown in Fig. \ref{fig:comparison:different:initial:condition}, we increase the initial phase angle of the first oscillator from 0~rad to 0.1~rad in the numerical simulations. In the second test, shown in Fig.~\ref{fig:comparison:different:manipulated:inputs}, we increase the manipulated inputs from $u = \begin{bmatrix} 1; 1 \end{bmatrix}$ to $u = \begin{bmatrix} 1.1; 1 \end{bmatrix}$.

In both tests, and for both the EN and the SM model, the balanced truncation approach 1) performs equally well for all tested values of $N$, and 2) performs as well or better than the POD approach.
Furthermore, when using the POD approach, some numerical simulations fail. This is not the case when using the balanced truncation approach.

% CREATE PLOTS OF L2 ERRORS (MODIFIED X0)

% CREATE PLOTS OF L2 ERRORS (MODIFIED U)
 % CHECK

\subsection{Reduction of the IEEE 118 bus system}\label{sec:test:118}
\begin{figure*}
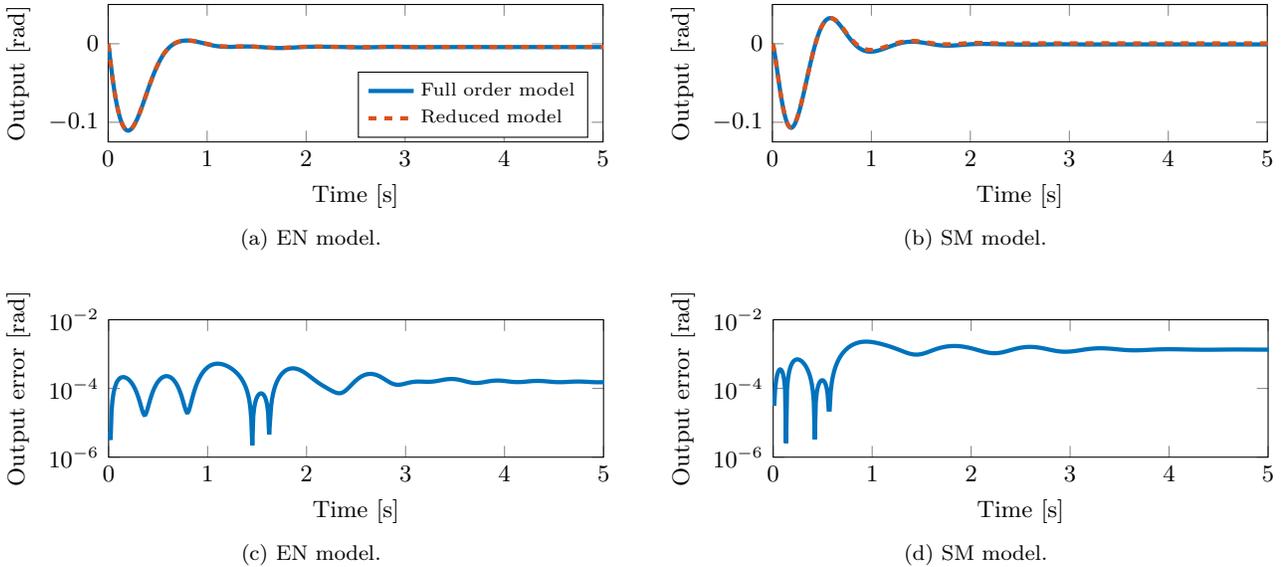

	\subfloat[EN model.\label{fig:case118_en}]{%
		\centering
		\legendenabledtrue % Enable legend
		\outputinitialconditionplot{case118}{en}
	}~
	\subfloat[SM model.\label{fig:case118_sm}]{%
		\centering
		\legendenabledfalse % Disable legend
		\outputinitialconditionplot{case118}{sm}
	}
	\\
	\subfloat[EN model.\label{fig:case118_error_en}]{%
		\centering
		\legendenabledtrue % Enable legend
		\outputerrorinitialconditionplot{case118}{en}
	}~
	\subfloat[SM model.\label{fig:case118_error_sm}]{%
		\centering
		\legendenabledfalse % Disable legend
		\outputerrorinitialconditionplot{case118}{sm}
	}
	\caption{Top row: Outputs for the original and the reduced EN and SM models of the IEEE 118 bus system. Bottom row: The absolute difference between the output for the original the reduced order models. The reduced models contain 20 oscillators, i.e., $n_r = 80$.}
	\label{fig:initial_condition:outputs}
\end{figure*}

Fig.~\ref{fig:initial_condition:outputs} shows the outputs and the output errors (as functions of time) for the original and the reduced EN and SM models of the IEEE 118 bus system. Based on the results in Section~\ref{sec:test:shift}~and~\ref{sec:compare:pod}, we use a shift of $\alpha = 5\cdot10^{-3}$ and $N = 3$ terms in the approximation of the Gramians in the balanced truncation approach.

Both of the reduced models contain 20 state variables, corresponding to a reduction of 63\% and 83\% for the EN and the SM model, respectively. 
Despite the large reductions, the outputs for the original and the reduced models are almost indistinguishable, and the absolute output errors do not exceed $10^{-3}$ for the EN model and $10^{-2}$ for the SM model.

% CREATE PLOTS
 % CHECK % CHECK
	\section{Conclusion}\label{sec:conclusions}
In this work, we describe a balanced truncation model reduction approach for reducing the nonlinear and dynamical EN and SM power grid models. In this approach, we 1)~reformulate the models as quadratic systems, 2)~approximate the Gramians of these systems, and 3)~use block-diagonal projection matrices in the balanced truncation.
We demonstrate the efficacy of this balanced truncation approach by reducing the IEEE~57 bus and IEEE~118 bus systems, and we compare it with a basic POD approach.

In future work, we will further investigate the relations between the choice of the projection matrices and 1)~the ability of the reduced EN and SM models to satisfy the PTI, $\sin^2(\delta_i) + \cos^2(\delta_i) = 1$, and 2)~their steady states. % CHECK
	
	% ACKNOWLEDGEMENTS
	\section*{Funding}
%\begin{funding}
	We acknowledge the financial support from the German Federal Ministry of Education and Research in the project KONSENS: Konsistente Optimierung und Stabilisierung Elektrischer Netzwerksysteme (BMBF grant 05M18EVA).
%\end{funding}
	
	\appendix
	
	% APPENDICES
	\section{Matricization}\label{sec:matricization}
We illustrate the concept of matricization using an example given by Kolda and Bader \cite{KolB09}. Let $\mathcal H\in\R^{3\times 4\times 2}$ be a tensor whose mode-1 matricization is
\begin{align}
	\mathcal H^{(1)} &=
	\begin{bmatrix}
		1 & 4 & 7 & 10 & 13 & 16 & 19 & 22 \\
		2 & 5 & 8 & 11 & 14 & 17 & 20 & 23 \\
		3 & 6 & 9 & 12 & 15 & 18 & 21 & 24
	\end{bmatrix}.
\end{align}
Then, the mode-2 and mode-3 matricizations are
\begin{subequations}
	\begin{align}
		\mathcal H^{(2)} &=
		\begin{bmatrix}
			 1 &  2 &  3 & 13 & 14 & 15 \\
			 4 &  5 &  6 & 16 & 17 & 18 \\
			 7 &  8 &  9 & 19 & 20 & 21 \\
			10 & 11 & 12 & 22 & 23 & 24
		\end{bmatrix}, \\
		\mathcal H^{(3)} &=
		\begin{bmatrix}
			 1 &  2 &  3 & \cdots & 10 & 11 & 12 \\
			13 & 14 & 15 & \cdots & 22 & 23 & 24
		\end{bmatrix}.
	\end{align}
\end{subequations}
For more information about matricization and tensors, we refer to \cite{KolB09} and to previous work on model reduction of quadratic-bilinear systems \cite{morBenB15, morBenG17}. % CHECK
	\section{Low-rank approximation}\label{sec:column:compression}
Given a matrix $R$ for which $P = R R^T$, we denote by $\tilde R = \mathcal T_\tau(R)$ a low-rank approximation based on the singular value decomposition, $R = U \Sigma V$:
\begin{align}
\tilde R &= U \Sigma_l.
\end{align}
$\Sigma_l$ contains the first $l$ columns of $\Sigma$, and $l$ is chosen such that
\begin{align}
\sigma_i^2 &> \tau \sigma_1^2,
\end{align}
for $i = 2, \ldots, l$.

In this work, we use the machine precision as the tolerance, i.e., $\tau = 1.1102\cdot10^{-16}$, in order to limit the error of the low-rank approximation. % CHECK
	\section{Efficient evaluation of the Kronecker products}\label{sec:efficient:evaluation:of:kronecker:products}
We evaluate $\Delta K_{ki}$ in \eqref{eq:kronecker:products:K} using matricization \cite{KolB09, morBenB15, morBenG17}:
\begin{subequations}
	\begin{align}
		\Delta K_{ik} &= \mathcal K_K^{(1)}, \\
		\mathcal K_K^{(3)} &= \tilde R_k \mathcal Y_K^{(3)}, \\
		\mathcal Y_K^{(2)} &= \tilde R_{i - (k+1)} H^{(2)}.
	\end{align}
\end{subequations}
Similarly, we evaluate $\Delta L_{ki}$ in \eqref{eq:kronecker:products:L} by
\begin{subequations}
	\begin{align}
		\Delta L_{ik} &= \mathcal K_L^{(2)}, \\
		\mathcal K_L^{(3)} &= \tilde R_k \mathcal Y_L^{(3)}, \\
		\mathcal Y_L^{(1)} &= \tilde S_{i - (k+1)} H.
	\end{align}
\end{subequations} % CHECK
	
	% REFERENCES
	\bibliographystyle{abbrvnat}
	\bibliography{./csc-bibfiles/csc.bib,./csc-bibfiles/mor.bib,./csc-bibfiles/software.bib,./ref/mor_power_systems.bib,./ref/misc.bib}

\begin{thebibliography}{8}
\providecommand{\natexlab}[1]{#1}
\providecommand{\url}[1]{\texttt{#1}}
\expandafter\ifx\csname urlstyle\endcsname\relax
  \providecommand{\doi}[1]{doi: #1}\else
  \providecommand{\doi}{doi: \begingroup \urlstyle{rm}\Url}\fi

\bibitem[Antoulas(2005)]{morAnt05}
A.~C. Antoulas.
\newblock \emph{Approximation of Large-Scale Dynamical Systems}, volume~6 of
  \emph{Adv. Des. Control}.
\newblock {SIAM} Publications, Philadelphia, PA, 2005.
\newblock ISBN 9780898715293.
\newblock \doi{10.1137/1.9780898718713}.

\bibitem[Benner and Breiten(2015)]{morBenB15}
P.~Benner and T.~Breiten.
\newblock Two-sided projection methods for nonlinear model order reduction.
\newblock \emph{{SIAM} J. Sci. Comput.}, 37\penalty0 (2):\penalty0 B239--B260,
  2015.
\newblock \doi{10.1137/14097255X}.

\bibitem[Benner and Goyal(2017)]{morBenG17}
P.~Benner and P.~Goyal.
\newblock Balanced truncation model order reduction for quadratic-bilinear
  systems.
\newblock e-prints 1705.00160, arXiv, 2017.
\newblock URL \url{https://arxiv.org/abs/1705.00160}.
\newblock math.OC.

\bibitem[Higham(1986)]{Hig86a}
N.~J. Higham.
\newblock Computing the polar decomposition---with applications.
\newblock \emph{{SIAM} J. Sci. Statist. Comput.}, 7:\penalty0 1160--1174, 1986.

\bibitem[Higham(1988)]{Hig88}
N.~J. Higham.
\newblock Computing a nearest symmetric positive semidefinite matrix.
\newblock \emph{Linear Algebra Appl.}, 103:\penalty0 103--118, 1988.
\newblock ISSN 0024-3795.
\newblock \doi{10.1016/0024-3795(88)90223-6}.

\bibitem[Horn and Johnson(1991)]{HorJ91}
R.~A. Horn and C.~R. Johnson.
\newblock \emph{Topics in Matrix Analysis}.
\newblock Cambridge University Press, Cambridge, 1991.

\bibitem[Nishikawa and Motter(2015)]{NisM15}
T.~Nishikawa and A.~E. Motter.
\newblock Comparative analysis of existing models for power-grid
  synchronization.
\newblock \emph{New J. Phys.}, 17:\penalty0 015012, Jan. 2015.
\newblock \doi{10.1088/1367-2630/17/1/015012}.

\bibitem[Schilders et~al.(2008)Schilders, {van der Vorst}, and
  Rommes]{morSchVR08}
W.~H.~A. Schilders, H.~A. {van der Vorst}, and J.~Rommes.
\newblock \emph{Model Order Reduction: Theory, Research Aspects and
  Applications}.
\newblock Springer-Verlag, Berlin, Heidelberg, 2008.

\end{thebibliography}


\begin{thebibliography}{56}
\providecommand{\natexlab}[1]{#1}
\providecommand{\url}[1]{\texttt{#1}}
\expandafter\ifx\csname urlstyle\endcsname\relax
  \providecommand{\doi}[1]{doi: #1}\else
  \providecommand{\doi}{doi: \begingroup \urlstyle{rm}\Url}\fi

\bibitem[Acle et~al.(2019)Acle, Freitas, Martins, and Rommes]{AclFMetal19}
Y.~G.~I. Acle, F.~D. Freitas, N.~Martins, and J.~Rommes.
\newblock Parameter preserving model order reduction of large sparse
  small-signal electromechanical stability power system models.
\newblock \emph{IEEE Transactions on Power Systems}, 34\penalty0 (4):\penalty0
  2814--2824, 2019.
\newblock \doi{10.1109/TPWRS.2019.2898977}.

\bibitem[Al-Saggaf(1993)]{Als93}
U.~M. Al-Saggaf.
\newblock Reduced-order models for dynamic control of a power plant with an
  improved transient and steady-state behavior.
\newblock \emph{Electric Power Systems Research}, 1993.

\bibitem[Antoulas(2005)]{morAnt05}
A.~C. Antoulas.
\newblock \emph{Approximation of Large-Scale Dynamical Systems}, volume~6 of
  \emph{Adv. Des. Control}.
\newblock {SIAM} Publications, Philadelphia, PA, 2005.
\newblock ISBN 9780898715293.
\newblock \doi{10.1137/1.9780898718713}.

\bibitem[Benner and Breiten(2015)]{morBenB15}
P.~Benner and T.~Breiten.
\newblock Two-sided projection methods for nonlinear model order reduction.
\newblock \emph{{SIAM} J. Sci. Comput.}, 37\penalty0 (2):\penalty0 B239--B260,
  2015.
\newblock \doi{10.1137/14097255X}.

\bibitem[Benner and Goyal(2017)]{morBenG17}
P.~Benner and P.~Goyal.
\newblock Balanced truncation model order reduction for quadratic-bilinear
  systems.
\newblock e-prints 1705.00160, arXiv, 2017.
\newblock URL \url{https://arxiv.org/abs/1705.00160}.
\newblock math.OC.

\bibitem[Chaniotis and Pai(2005)]{ChaP05}
D.~Chaniotis and M.~A. Pai.
\newblock Model reduction in power systems using {K}rylov subspace methods.
\newblock \emph{IEEE Transactions on Power Systems}, 20\penalty0 (2):\penalty0
  888--894, 2005.
\newblock \doi{10.1109/TPWRS.2005.846109}.

\bibitem[Cheng and Scherpen(2018)]{morCheS18}
X.~Cheng and J.~M.~A. Scherpen.
\newblock Clustering approach to model order reduction of power networks with
  distributed controllers.
\newblock \emph{Advances in Computational Mathematics}, 44\penalty0
  (6):\penalty0 1917--1939, 2018.
\newblock \doi{10.1007/s10444-018-9617-5}.

\bibitem[Cherid and Bettayeb(1991)]{CheM91}
A.~Cherid and M.~Bettayeb.
\newblock Reduced-order models for the dynamics of a single-machine power
  system via balancing.
\newblock \emph{Electric Power Systems Research}, 22\penalty0 (1):\penalty0
  7--12, 1991.
\newblock \doi{10.1016/0378-7796(91)90073-V}.

\bibitem[Chow(2013)]{Cho13}
J.~H. Chow, editor.
\newblock \emph{Power system coherency and model reduction}, volume~94.
\newblock Springer, 2013.
\newblock \doi{10.1007/978-1-4614-1803-0}.

\bibitem[Chow et~al.(1990)Chow, Winkelman, Pai, and Sauer]{ChoWPetal90}
J.~H. Chow, J.~R. Winkelman, M.~A. Pai, and P.~W. Sauer.
\newblock Singular perturbation analysis of large-scale power systems.
\newblock \emph{International Journal of Electrical Power \& Energy Systems},
  12\penalty0 (2), 1990.
\newblock \doi{10.1016/0142-0615(90)90007-X}.

\bibitem[Golub and {Van~Loan}(2013)]{GolV13}
G.~H. Golub and C.~F. {Van~Loan}.
\newblock \emph{Matrix Computations}.
\newblock Johns Hopkins Studies in the Mathematical Sciences. Johns Hopkins
  University Press, Baltimore, fourth edition, 2013.
\newblock ISBN 978-1-4214-0794-4; 1-4214-0794-9; 978-1-4214-0859-0.

\bibitem[Higham(1986)]{Hig86a}
N.~J. Higham.
\newblock Computing the polar decomposition---with applications.
\newblock \emph{{SIAM} J. Sci. Statist. Comput.}, 7:\penalty0 1160--1174, 1986.

\bibitem[Higham(1988)]{Hig88}
N.~J. Higham.
\newblock Computing a nearest symmetric positive semidefinite matrix.
\newblock \emph{Linear Algebra Appl.}, 103:\penalty0 103--118, 1988.
\newblock ISSN 0024-3795.
\newblock \doi{10.1016/0024-3795(88)90223-6}.

\bibitem[Hockenberry(2000)]{Hoc2000}
J.~R. Hockenberry.
\newblock \emph{Evaluation of uncertainty in dynamic reduced-order power system
  models}.
\newblock PhD thesis, Massachusetts Institute of Technology, 2000.

\bibitem[Horn and Johnson(1991)]{HorJ91}
R.~A. Horn and C.~R. Johnson.
\newblock \emph{Topics in Matrix Analysis}.
\newblock Cambridge University Press, Cambridge, 1991.

\bibitem[Joo et~al.(2004)Joo, Liu, Jones, and Choe]{JooLJ04}
S.-K. Joo, C.-C. Liu, L.~E. Jones, and J.-W. Choe.
\newblock Coherency and aggregation techniques incorporating rotor and voltage
  dynamics.
\newblock \emph{IEEE Transactions on Power Systems}, 19\penalty0 (2):\penalty0
  1068--1075, 2004.
\newblock \doi{10.1109/TPWRS.2004.825825}.

\bibitem[Khatibi and Zargarzadeh(2016)]{KhaZ16}
M.~Khatibi and H.~Zargarzadeh.
\newblock Power system dynamic model reduction by means of an iterative
  {SVD}-{K}rylov model reduction method.
\newblock In \emph{2016 IEEE Power \& Energy Society Innovative Smart Grid
  Technologies Conference (ISGT)}, page 16526042, 2016.
\newblock \doi{10.1109/ISGT.2016.7781027}.

\bibitem[Kolda and Bader(2009)]{KolB09}
T.~G. Kolda and B.~W. Bader.
\newblock Tensor decompositions and applications.
\newblock \emph{{SIAM} Rev.}, 51\penalty0 (3):\penalty0 455--500, 2009.
\newblock ISSN 0036-1445.
\newblock \doi{10.1137/07070111X}.

\bibitem[Kramer and Willcox(2019{\natexlab{a}})]{KraW19b}
B.~Kramer and K.~Willcox.
\newblock Balanced truncation model reduction for lifted nonlinear systems,
  2019{\natexlab{a}}.
\newblock arXiv: 1907.12084v1.

\bibitem[Kramer and Willcox(2019{\natexlab{b}})]{KraW19}
B.~Kramer and K.~E. Willcox.
\newblock Nonlinear model order reduction via lifting transformations and
  proper orthogonal decomposition.
\newblock \emph{AIAA Journal}, 57\penalty0 (6):\penalty0 2297--2307,
  2019{\natexlab{b}}.
\newblock \doi{10.2514/1.J057791}.

\bibitem[Lan et~al.(2016)Lan, Zhao, Wang, and Mi]{LanZWetal16}
X.~Lan, H.~Zhao, Y.~Wang, and Z.~Mi.
\newblock Nonlinear power system model reduction based on empirical {G}ramians.
\newblock In \emph{2016 IEEE International Conference on Power System
  Technology (POWERCON)}, page 16487940, 2016.
\newblock \doi{10.1109/POWERCON.2016.7754074}.

\bibitem[Leung et~al.(2019)Leung, Kinnaert, Maun, and Villella]{LeuKMetal19b}
J.~Leung, M.~Kinnaert, J.-C. Maun, and F.~Villella.
\newblock Model reduction of coherent {LPV} models in power systems.
\newblock In \emph{2019 IEEE Power \& Energy Society General Meeting (PESGM)},
  page 19302246, 2019.
\newblock \doi{10.1109/PESGM40551.2019.8973651}.

\bibitem[Levron and Belikov(2017)]{LevB17}
Y.~Levron and J.~Belikov.
\newblock Reduction of power system dynamic models using sparse
  representations.
\newblock \emph{IEEE Transactions on Power Systems}, 32\penalty0 (5):\penalty0
  3893--3900, 2017.
\newblock \doi{10.1109/TPWRS.2017.2648979}.

\bibitem[Malik et~al.(2016)Malik, Borzacchiello, Chinesta, and
  Diez]{MalBCetal16}
M.~H. Malik, D.~Borzacchiello, F.~Chinesta, and P.~Diez.
\newblock Reduced order modeling for transient simulation of power systems
  using tracetory piece-wise linear approximation.
\newblock \emph{Advanced Modeling and Simulation in Engineering Sciences},
  3:\penalty0 31, 2016.
\newblock \doi{10.1186/s40323-016-0084-6}.

\bibitem[Meng et~al.(2018)Meng, Wang, Zhou, Xiao, and Chi]{MenWZetal18}
X.~Meng, Q.~Wang, N.~Zhou, S.~Xiao, and Y.~Chi.
\newblock Multi-time scale model order reduction and stability consistency
  certification of inverter-interfaced {DG} system in {AC} microgrid.
\newblock \emph{Energies}, 11\penalty0 (1):\penalty0 254, 2018.
\newblock \doi{10.3390/en11010254}.

\bibitem[Milano and Srivastava(2009)]{MilS09}
F.~Milano and K.~Srivastava.
\newblock Dynamic {REI} equivalents for short circuit and transient stability
  analyses.
\newblock \emph{Electric Power Systems Research}, 79\penalty0 (6):\penalty0
  878--887, 2009.
\newblock \doi{10.1016/j.epsr.2008.11.007}.

\bibitem[Mlinari\'{c} et~al.(2018)Mlinari\'{c}, Ishizaki, Chakrabortty,
  Grundel, Benner, and Imura]{MliICetal18}
P.~Mlinari\'{c}, T.~Ishizaki, A.~Chakrabortty, S.~Grundel, P.~Benner, and J.-i.
  Imura.
\newblock Synchronization and aggregation of nonlinear power systems with
  consideration of bus network structures.
\newblock In \emph{2018 European Control Conference (ECC)}, pages 2266--2271,
  2018.
\newblock \doi{10.23919/ECC.2018.8550528}.

\bibitem[Nishikawa and Motter(2015)]{NisM15}
T.~Nishikawa and A.~E. Motter.
\newblock Comparative analysis of existing models for power-grid
  synchronization.
\newblock \emph{New J. Phys.}, 17:\penalty0 015012, Jan. 2015.
\newblock \doi{10.1088/1367-2630/17/1/015012}.

\bibitem[Osipov and Sun(2018)]{OsiS18}
D.~Osipov and K.~Sun.
\newblock Adaptive nonlinear model reduction for fast power system simulation.
\newblock \emph{IEEE Transactions on Power Systems}, 33\penalty0 (6):\penalty0
  6746--6754, 2018.
\newblock \doi{10.1109/TPWRS.2018.2835766}.

\bibitem[Osipov et~al.(2018)Osipov, Duan, Dimitrovski, Allu, Simunovic, and
  Sun]{OsiDDetal18}
D.~Osipov, N.~Duan, A.~Dimitrovski, S.~Allu, S.~Simunovic, and K.~Sun.
\newblock Adaptive model reduction for parareal in time method for transient
  stability simulations.
\newblock In \emph{2018 IEEE Power \& Energy Society General Meeting (PESGM)},
  2018.

\bibitem[Pai and Adgaonkar(1981)]{PaiA81}
M.~A. Pai and R.~P. Adgaonkar.
\newblock Singular perturbation analysis of nonlinear transients in power
  systems.
\newblock In \emph{1981 20th IEEE Conference on Decision and Control including
  the Symposium on Adaptive Processes}, pages 221--222, 1981.
\newblock \doi{10.1109/CDC.1981.269515}.

\bibitem[Parang et~al.(2019)Parang, Mohammadi, and Arefi]{ParMA19}
B.~Parang, M.~Mohammadi, and M.~M. Arefi.
\newblock Residualisation-based model order reduction in power networks with
  penetration of photovoltaic resources.
\newblock \emph{IET Generation, Transmission \& Distribution}, 13\penalty0
  (13):\penalty0 2619--2626, 2019.
\newblock \doi{10.1049/iet-gtd.2018.6172}.

\bibitem[Parrilo et~al.(1999)Parrilo, Lall, Paganini, Verghese, Lesieutre, and
  Marsden]{ParLPetal99}
P.~A. Parrilo, S.~Lall, F.~Paganini, G.~C. Verghese, B.~C. Lesieutre, and J.~E.
  Marsden.
\newblock Model reduction for analysis of cascading failures in power systems.
\newblock In \emph{1999 American Control Conference (ACC)}, 1999.
\newblock \doi{10.1109/ACC.1999.786351}.

\bibitem[Purvine et~al.(2017)Purvine, Cotilla-Sanchez, Halappanavar, Huang,
  Lin, Lu, and Wang]{PurCHetal17}
E.~Purvine, E.~Cotilla-Sanchez, M.~Halappanavar, Z.~Huang, G.~Lin, S.~Lu, and
  S.~Wang.
\newblock Comparative study of clustering techniques for real-time dynamic
  model reduction.
\newblock \emph{Statistical Analysis and Data Mining}, 10\penalty0
  (5):\penalty0 263--276, 2017.
\newblock \doi{10.1002/sam.11352}.

\bibitem[Ramirez et~al.(2016)Ramirez, Mehrizi-Sani, Hussein, Matar,
  Abdel-Rahman, Chavez, Davoudi, and Kamalasadan]{RamMHetal16}
A.~Ramirez, A.~Mehrizi-Sani, D.~Hussein, M.~Matar, M.~Abdel-Rahman, J.~J.
  Chavez, A.~Davoudi, and S.~Kamalasadan.
\newblock Application of balanced realizations for model-order reduction of
  dynamic power system equivalents.
\newblock \emph{IEEE Transactions on Power Delivery}, 31\penalty0 (5):\penalty0
  2304--2312, 2016.
\newblock \doi{10.1109/TPWRD.2015.2496498}.

\bibitem[Rudnick et~al.(1981)Rudnick, Patino, and Brameller]{RudPB81}
H.~Rudnick, R.~I. Patino, and A.~Brameller.
\newblock Power-system dynamic equivalents: coherency recognition via the rate
  of change of kinetic energy.
\newblock \emph{IEE Proceedings C - Generation, Transmission and Distribution},
  pages 325--333, 1981.
\newblock \doi{10.1049/ip-c.1981.0052}.

\bibitem[Sanchez-Gasca et~al.(1995)Sanchez-Gasca, Chow, and Galarza]{SanCG95}
J.~J. Sanchez-Gasca, J.~H. Chow, and R.~Galarza.
\newblock Reduction of linearized power systems for the study of interarea
  oscillations.
\newblock In \emph{International Conference on Control Applications (CCA)},
  page 5119382, 1995.
\newblock \doi{10.1109/CCA.1995.555806}.

\bibitem[Scarciotti(2015)]{Sca15}
G.~Scarciotti.
\newblock Model reduction of power systems with preservation of slow and poorly
  damped modes.
\newblock In \emph{2015 IEEE Power \& Energy Society General Meeting}, page
  15502166, 2015.
\newblock \doi{10.1109/PESGM.2015.7285719}.

\bibitem[Schilders et~al.(2008)Schilders, {van der Vorst}, and
  Rommes]{morSchVR08}
W.~H.~A. Schilders, H.~A. {van der Vorst}, and J.~Rommes.
\newblock \emph{Model Order Reduction: Theory, Research Aspects and
  Applications}.
\newblock Springer-Verlag, Berlin, Heidelberg, 2008.

\bibitem[Shomalzadeh and Amraee(2017)]{ShoA17}
K.~Shomalzadeh and T.~Amraee.
\newblock Unstable power system model reduction using balanced truncation.
\newblock In \emph{2017 25th Iranian Conference on Electrical Engineering
  (ICEE)}, page 17045511, 2017.
\newblock \doi{10.1109/IranianCEE.2017.7985241}.

\bibitem[Sturk(2012)]{Stu12}
C.~Sturk.
\newblock \emph{Structured model reduction and its application to power
  systems}.
\newblock PhD thesis, KTH Royal Institute of Technology, 2012.

\bibitem[Sturk et~al.(2012{\natexlab{a}})Sturk, Vanfretti, Chompoobutrgool, and
  Sandberg]{StuVCetal12}
C.~Sturk, L.~Vanfretti, Y.~Chompoobutrgool, and H.~Sandberg.
\newblock Structured power system model reduction of non-coherent areas.
\newblock In \emph{2012 IEEE Power and Energy Society General Meeting (PESGM)},
  page 13170255, 2012{\natexlab{a}}.
\newblock \doi{10.1109/PESGM.2012.6344913}.

\bibitem[Sturk et~al.(2012{\natexlab{b}})Sturk, Vanfretti, Milano, and
  Sandberg]{StuVMetal12}
C.~Sturk, L.~Vanfretti, F.~Milano, and H.~Sandberg.
\newblock Structured model reduction of power systems.
\newblock In \emph{2012 American Control Conference (ACC)}, pages 2276--2282,
  2012{\natexlab{b}}.
\newblock \doi{10.1109/ACC.2012.6315207}.

\bibitem[Sturk et~al.(2014)Sturk, Vanfretti, Chompoobutrgool, and
  Sandberg]{StuVCetal14}
C.~Sturk, L.~Vanfretti, Y.~Chompoobutrgool, and H.~Sandberg.
\newblock Coherency-independent structured model reduction of power systems.
\newblock \emph{IEEE Transactions on Power Systems}, 29\penalty0 (5):\penalty0
  2418--2426, 2014.
\newblock \doi{10.1109/TPWRS.2014.2302871}.

\bibitem[{Van Loan}(2000)]{Van00}
C.~F. {Van Loan}.
\newblock The ubiquitous {K}ronecker product.
\newblock \emph{Journal of Computational and Applied Mathematics},
  123:\penalty0 85--100, 2000.
\newblock \doi{10.1016/S0377-0427(00)00393-9}.

\bibitem[Wang et~al.(2013)Wang, Yu, Li, Ding, Sun, Guo, Zhang, Zhou, and
  Yu]{WanYLetal13}
C.~Wang, H.~Yu, P.~Li, C.~Ding, C.~Sun, X.~Guo, F.~Zhang, Y.~Zhou, and Z.~Yu.
\newblock {K}rylov subspace based model reduction method for transient
  simulation of active distribution grid.
\newblock In \emph{2013 IEEE Power \& Energy Society General Meeting}, page
  13933173, 2013.
\newblock \doi{10.1109/PESMG.2013.6672277}.

\bibitem[Wang et~al.(2015)Wang, Yu, Li, Wu, and Ding]{WanYLetal15}
C.~Wang, H.~Yu, P.~Li, J.~Wu, and C.~Ding.
\newblock Model order reduction for transient simulation of active distribution
  networks.
\newblock \emph{IET Generation, Transmission \& Distribution}, 9\penalty0 (5),
  2015.
\newblock \doi{10.1049/iet-gtd.2014.0219}.

\bibitem[Wang et~al.(2014)Wang, Lu, Zhou, Lin, Elizondo, and Pai]{WanLZetal14}
S.~Wang, S.~Lu, N.~Zhou, G.~Lin, M.~Elizondo, and M.~A. Pai.
\newblock Dynamic-feature extraction, attribution, and reconstruction ({DEAR})
  method for power system model reduction.
\newblock \emph{IEEE Transactions on Power Systems}, 29\penalty0 (5):\penalty0
  2049--2059, 2014.
\newblock \doi{10.1109/TPWRS.2014.2301032}.

\bibitem[Wang et~al.(2006)Wang, Huang, and Wu]{WanHW06}
S.-C. Wang, P.-H. Huang, and C.-J. Wu.
\newblock Application of fuzzy {C}-means clustering in power system model
  reduction for controller design.
\newblock In \emph{5th WSEAS International Conference on Computational
  Intelligence, Man-machine systems and Cybernetics}, pages 223--227, 2006.

\bibitem[Weber and Welfonder(1988)]{WebW88}
H.~Weber and E.~Welfonder.
\newblock Dynamic model reduction for the modal analysis of frequency and power
  oscillations in large power systems.
\newblock \emph{IFAC Proceedings Volumes}, 21\penalty0 (11):\penalty0 233--239,
  1988.
\newblock \doi{10.1016/S1474-6670(17)53749-0}.

\bibitem[Wei{\ss}(2019)]{Wei19}
F.~Wei{\ss}.
\newblock Simulation, analysis, and model order reduction for dynamic power
  network models.
\newblock Master's thesis, Otto-von-Guericke-Universit{\"{a}}t Magdeburg, 2019.

\bibitem[Zhao et~al.(2017)Zhao, Lan, and Ren]{ZhaLR17}
H.~Zhao, X.~Lan, and H.~Ren.
\newblock Nonlinear power system model reduction based on empirical {G}ramians.
\newblock \emph{Journal of Electrical Engineering}, 68\penalty0 (6):\penalty0
  425--434, 2017.
\newblock \doi{10.1515/jee-2017-0077}.

\bibitem[Zhao et~al.(2013)Zhao, Xue, and Shi]{ZhaXS13}
H.-S. Zhao, N.~Xue, and N.~Shi.
\newblock Nonlinear dynamic power system model reduction analysis using
  balanced empirical {G}ramian.
\newblock \emph{Applied Mechanics and Materials}, 448--453:\penalty0
  2368--2374, 2013.
\newblock \doi{10.4028/www.scientific.net/AMM.448-453.2368}.

\bibitem[Zhu et~al.(2016)Zhu, Geng, and Jiang]{ZhuGQ16}
Z.~Zhu, G.~Geng, and Q.~Jiang.
\newblock Power system dynamic model reduction based on extended {K}rylov
  subspace method.
\newblock \emph{IEEE Transactions on Power Systems}, 31\penalty0 (6):\penalty0
  4483--4494, 2016.
\newblock \doi{10.1109/TPWRS.2015.2509481}.

\bibitem[Zimmerman and Murillo-S{\'{a}}nchez(2016)]{ZimM16}
R.~D. Zimmerman and C.~E. Murillo-S{\'{a}}nchez.
\newblock {MATPOWER} (version 6.0), 2016.

\bibitem[Zimmerman et~al.(2011)Zimmerman, Murillo-S{\'{a}}nchez, and
  Thomas]{ZimMT11}
R.~D. Zimmerman, C.~E. Murillo-S{\'{a}}nchez, and R.~J. Thomas.
\newblock {MATPOWER}: Steady-state operations, planning, and analysis tools for
  power systems research and education.
\newblock \emph{IEEE Transactions on Power Systems}, 26\penalty0 (1):\penalty0
  12--19, 2011.
\newblock \doi{10.1109/TPWRS.2010.2051168}.

\end{thebibliography}
\end{document}